# Telling Great Stories: An NSDL Content and Communications System for Aggregation, Display, and Distribution of News and Features

Carol Minton Morris (Terrizzi)
Cornell Information Science
301 College Avenue
Ithaca, New York 14850
607 255-2702

Clt6@cornell.edu

## Abstract
Education digital libraries contain cataloged resources as well as contextual information about innovations in the use of educational technology, exemplar stories about community activities, and news from various user communities that include teachers, students, scholars, and developers. Long-standing library traditions of service, preservation, democratization of knowledge, rich discourse, equal access, and fair use are evident in library communications models that both pull in and push out contextual information from multiple sources integrated with editorial production processes. This paper argues that a dynamic narrative flow [1] is enabled by effective management of complex content and communications in a decentralized web-based education digital library making publishing objects such as aggregations of resources, or selected parts of objects [4] accessible through a Content and Communications System. Providing services that encourage patrons to reuse, reflect out, and contribute resources back [5] to the Library increases the reach and impact of the National Science Digital Library (NSDL). This system is a model for distributed content development and effective communications for education digital libraries in general.

## Introduction
To become what NSDL Director Kaye Howe has called, "A center of information and leadership on digital libraries and education," an NSDL Content and Communications System with embedded user tools for discovery and reuse enables leveraged public awareness[1] and dissemination of publishing objects whose origin is in the digital library.

**ACKNOWLEDGEMENT**
This material is based upon work supported by the National Science Foundation under Grants No. 0227648, 0227656, and 0227888. Any opinions, findings, and conclusions or recommendations expressed in this material are those of the author and do not necessarily reflect the views of the National Science Foundation.

New publishing objects will result by combining an array of library content limited only by users' imaginations in telling great STEM learning stories across multiple environments.

## Previous Work
NSDL communications activities seek to ensure that NSDL's community of collaborators who are engaged in using, building, and promoting the digital library see themselves as contributors to, and beneficiaries of a rich information exchange among diverse stakeholders [3]. Flexible content that is small, modular, and adaptable is favored in creating a reusable and multilayered flow of information in an educational digital library. Collaborative document development is preferable to a top-down approach to publishing. [4]

In collaboration with The Internet Scout Project at the University of Wisconsin, NSDL Core Integration developed a content management system based on Scout Portal Toolkit technology as a model for an interactive NSDL Content and Communications System. This metadata-based system pulls in, aggregates, and publishes, or pushes out, searchable text and image news and features publishing objects. The system relies on non-automated, centralized, human interaction with content databases to manipulate content.

## Related Work
Fedora is a general-purpose digital object repository system that can be used in whole or part to support a variety of uses including: institutional repositories, digital libraries, content management, digital asset management, scholarly publishing, and digital preservation. [4] The NSDL Fedora Repository is being developed by the Cornell University Information Science Program.

The proposed NSDL Content and Communications System utilizes the NSDL Fedora repository's unique

---

[1] Example: NSF's NSDL "Classroom Resources" display at: http://nsf.gov/news/classroom/.

**Figure 1. NSDL Content and Communications System.**

**Figure 2. Submitting a news item through the NSDL Content and Communications System.**

characteristics with respect to maintaining relationships and applications [7] associated with complex digital objects in a user interface (Fig. 1), ultimately disseminating multiple types of news and features objects to internal and external audiences [2].

## Content Types

Value-added contextual information (Fig. 1) highlights NSDL resources for library patrons in the same way that exhibits in physical libraries display books, objects, and event information in glass cases. NSDL content falls into two categories: news or features. News exhibits may be combinations of text, images, media, and links that are time-sensitive, brief, and structurally simple. Features exhibits differ in that they are not time-sensitive, may be longer in length, and include a more complex document structure that augments the meaning of the content by calling attention to parts of the document. News and features content types in Figure 1 have been analyzed for design of storage, retrieval, and dissemination of content data streams via Fedora.

## Flexible Publishing Object-Based Content and Communications

In a dynamically-driven communications paradigm "containers" for publishing such as newsletters, magazines, RSS feeds, books, blogs, brochures, or reports are less important than individual stories, news items, and images available through the library for contextual re-packaging and reflecting out (Fig. 2). NSDL's interactive Content and Communications System enables a democratic narrative [6] based on randomness, metadata, and dynamism [1] in user interactions with the NSDL Repository.

## Conclusion

New content types will emerge from a user-oriented NSDL Content and Communications System that tells the story of The National Science Digital Library as users integrate rich and varied STEM information based on their own criteria to discover and communicate knowledge.

## Acknowledgements

The author would like to extend special thanks to Kaye Howe, Director, NSDL, Susan Jesuroga, Director, NSDL-Funded Project Relations, Len Simutis, Director, Eisenhower National Clearinghouse (ENC), Rachael Bower, Co-Director, The Internet Scout Project, Paul Berkman, University of California, Santa Barbara, Ellen J. Cramer, NSDL Systems Analyst and Developer, and the NSDL FEDORA team: Carl Lagoze, Sandy Payette, Ellen Cramer, Edwin Shin, Chris Wilper, and Dean Krafft.